\documentstyle[twoside,fleqn,espcrc2,epsfig]{article}

\newcommand{\AmS}{{\protect\the\textfont2
  A\kern-.1667em\lower.5ex\hbox{M}\kern-.125emS}}

\def\kreis{\raise0.85pt\hbox{$\scriptscriptstyle\bigcirc$}}
\def\vollk{\lower0.95pt\hbox{\Large $\bullet$}}
\def\diamo{\lower0.45pt\hbox{$\Diamond$}}

\title{
\vspace{-2.5cm}                         
{\normalsize DESY 98--146}     \\[-0.2cm]      
{\normalsize FU-HEP/98--5}    \\[-0.2cm]      
{\normalsize HUB--EP--98/56}   \\[-0.2cm]      
{\normalsize TPR--98--26}   \\[-0.2cm]      
{\normalsize HLRZ 1998--57}   \\[-0.2cm]      
{\normalsize September 1998}      \\              
\vspace{0.7cm} 
Resolving Exceptional Configurations in Quenched 
Lattice QCD\thanks{Poster presented by G. Schierholz at Lattice 1998.}}

\author{
        M.~G\"ockeler$^{\rm a}$, 
    A.~Hoferichter$^{\rm b}$,
    R.~Horsley$^{\rm c}$, 
    D.~Pleiter$^{\rm b,d}$,
    P.~Rakow$^{\rm a}$,
    G.~Schierholz$^{\rm b,e}$,\\
    and P.~Stephenson$^{\rm f}$ \\[1em]
    $^{\rm a}$ Institut f\"ur Theoretische Physik, 
    Universit\"at Regensburg, D-93040 Regensburg \\[0.5em]
    $^{\rm b}$ Deutsches Elektronen-Synchrotron DESY and NIC, D-15735 
    Zeuthen\\[0.5em]
    $^{\rm c}$ Institut f\"ur Physik, Humboldt-Universit\"at zu Berlin, 
    D-10115 Berlin\\[0.5em]
    $^{\rm d}$ Institut f\"ur Theoretische Physik, Freie Universit\"at
    Berlin, D-14195 Berlin\\[0.5em]
    $^{\rm e}$ Deutsches Elektronen-Synchrotron DESY, 
    D-22603 Hamburg\\[0.5em]
    $^{\rm f}$ Dipartimento di Fisica, Universit\`a di Pisa,
    I-56126 Pisa\\[0.5em]}
       
\begin{document}

\begin{abstract}
Quenched lattice QCD calculations with Wilson-type fermions at small quark
masses are impeded by exceptional configurations. We show how this
problem can be resolved in a practicable way without changing the
physics in the chiral limit.
\end{abstract}

\maketitle

\setcounter{footnote}{0}

\section{INTRODUCTION}
\label{intro}

We consider Wilson fermions with or without $O(a)$ improvement. The
fermionic action reads
\begin{equation}
S_F = \sum \bar{\psi}{\cal K}\psi, \; {\cal K} = W + M,
\label{action}
\end{equation}
where $W$ is the Wilson-Dirac operator:
\begin{equation}
W = \not{\!\!D} + X
\end{equation} 
with
\begin{equation}
D_\mu = \frac{1}{2}(D^+_\mu + D^-_\mu), 
\end{equation}
\vspace*{-0.3cm}
\begin{equation}
X = - \frac{1}{2} D^+_\mu D^-_\mu 
 + \frac{\mbox{i}}{4}c_{\scriptstyle {\rm SW}}\, 
\sigma_{\mu\nu}F_{\mu\nu}, 
\end{equation}
$D^+_\mu$ ($D^-_\mu$) being the forward (backward) lattice
derivative, and where $M$ is the Wilson mass term:
\begin{equation}
M = \frac{1}{2\kappa} - 4.
\end{equation}
The critical value of the hopping parameter, $\kappa_c$,
is taken to be the limiting value at which the ensemble averaged pion mass
vanishes. We call this limit the chiral limit.

\begin{figure}
\vspace{-0.5cm}
\begin{centering}
\epsfig{figure=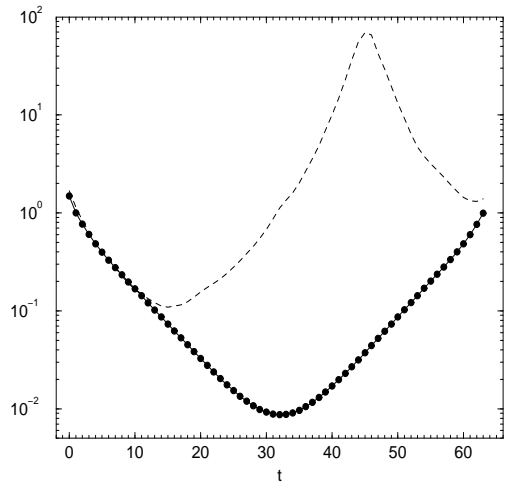,height=6.5cm,width=6.5cm}
\vspace{-0.7cm}
\caption{The pion correlator on the $32^3 64$ lattice at $\beta = 6.2$
  and $\kappa = 0.1354$ for improved fermions with
  $c_{\scriptstyle {\rm SW}} = 1.614$. At these parameters
  $m_\pi/m_\rho \approx 0.5$ which corresponds to a quark mass of
  $\approx 40 \mbox{MeV}$. The solid symbols (and
  line) represent the ensemble average with the exceptional
  configurations omitted. The dashed line is the result of a single,
  particularly exceptional configuration.}
\vspace{-0.5cm}
\end{centering}
\end{figure}

\begin{figure*}[bht]
\vspace{-0.35cm}
\begin{centering}
\epsfig{figure=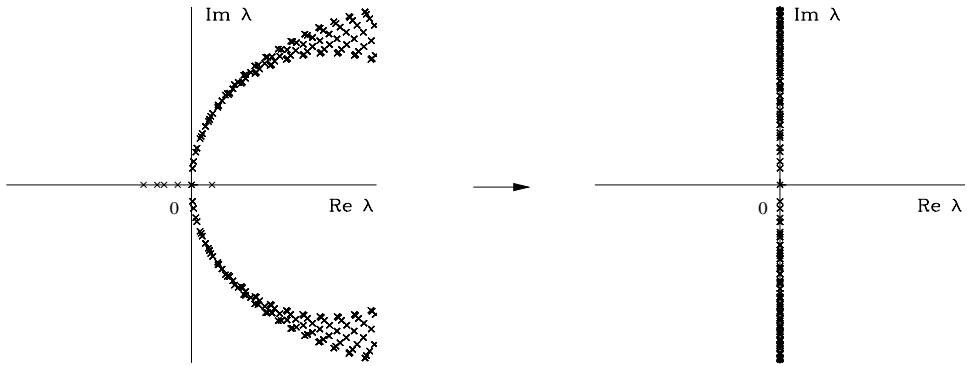,height=5.75cm,width=13.5cm,bbllx=63,bblly=286,
bburx=560,bbury=506}
\vspace{-0.7cm}
\caption{A sketch of the eigenvalue spectrum of $W_c$ (before the
  rotation) and $W_c^{\phi = \pi/4}$ (after the rotation).}
\end{centering}
\end{figure*}

In quenched lattice calculations with Wilson fermions one encounters
exceptional configurations~\cite{exceptional} which give
rise to poles in fermionic observables, such as hadron correlators, at
small quark masses. The origin of these poles are real eigenvalues  
$\lambda_i$ of the Wilson-Dirac operator $W$ at 
\begin{equation}
\lambda_i < 4 - \frac{1}{2 \kappa_c}.
\label{eig}
\end{equation} 
Exceptional configurations are particularly aggravating for improved
fermions, and they have limited calculations to larger quark masses
and larger values of $\beta$. 
We expect to find such poles also in the full theory, i.e. with 
dynamical fermions, at hopping parameters $\kappa \neq
\kappa_{\scriptstyle {\rm sea}}$, where
$\kappa_{\scriptstyle {\rm sea}}$ refers to the sea quark mass.

In Fig.~1 we give an example of an exceptional configuration which
we encountered in a recent simulation with improved
fermions~\cite{Dirk}. We see a dramatic deflection from the anticipated
ensemble average. Similar peaks are observed in the $\rho$ and nucleon
correlators. It is quite clear from this figure that the problem of
exceptional configurations cannot be
solved by accumulating larger statistical samples.  

Bardeen et al.~\cite{Bardeen} identify the eigenvalues (\ref{eig}) 
with the zero modes of the continuum theory, and they propose to shift each 
such $\lambda_i$ to $\lambda_i = 4 - 1/2 \kappa_c$. There are doubts, however, 
that this identification is correct in general~\cite{Gattringer}. 
An elegant solution would be to employ Neuberger's action~\cite{Neuberger}.
But whether this is numerically feasible has still to be seen.

\section{PROPOSAL}

The solution we propose makes use of a freedom in regularizing lattice
fermions. It has the advantage that it is easy to implement numerically, and
that it does not change the physics in the chiral limit so that the 
improvement program \`{a} la Sheikholeslami and Wohlert~\cite{Jansen}
remains unaltered.

We rewrite ${\cal K}$ in (\ref{action}) as
\begin{equation}
{\cal K} = W_c + m,
\label{caction}
\end{equation}
where $W_c$ and $m$ are the critical Wilson-Dirac operator and the
physical bare mass, respectively:
\begin{equation}
W_c = \not{\!\!D} + X_c,\; X_c = X + \frac{1}{2\kappa_c} - 4,
\end{equation}
\vspace*{-0.4cm}
\begin{equation} 
m = \frac{1}{2\kappa} - \frac{1}{2\kappa_c}.
\end{equation}
We then apply a chiral rotation to $W_c$,
\begin{eqnarray}
W_c \rightarrow W_c^\phi \!\!\!&=&\!\!\! 
\mbox{e}^{{\scriptstyle \rm i}\gamma_5\phi} W_c 
\mbox{e}^{{\scriptstyle \rm i} \gamma_5\phi}, \nonumber \\
\!\!\!&=&\!\!\! \not{\!\!D} + X_c \,(\cos 2\phi + \mbox{i}\gamma_5 \sin
2\phi),
\label{transf}
\end{eqnarray}
and take ${\cal K}^\phi \equiv W_c^\phi + m$ to be the new fermion matrix.
The resulting action is equally well founded as
(\ref{action}) and (\ref{caction})~\cite{I&Y}, though, in general, 
it breaks parity invariance. In Fig.~2 we
show the effect of the transformation on the eigenvalue spectrum of
the Wilson-Dirac operator for $\phi = \pi/4$. In this case $W_c$ is
anti-hermitian, and all eigenvalues are
mapped onto the imaginary axis, as is the case for staggered fermions.

\begin{figure}
\begin{centering}
\hspace*{-0.5cm} \epsfig{figure=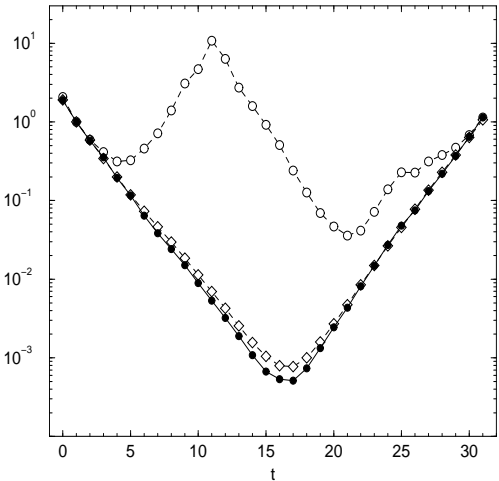,height=6.5cm,width=6.5cm}
\vspace{-0.7cm}
\caption{The pion correlator for a single exceptional configuration on
  the $16^3 32$ lattice at $\beta = 5.7$ and 
  $c_{\scriptstyle {\rm SW}} = 2.25, \kappa = 0.1295$ before
  ($\kreis$) and after the rotation by $\phi = \pi/4$ ($\vollk$),
  with $\kappa_c = 0.13074$~\cite{Gockeler}. The data for improved 
  fermions are compared with the result for Wilson fermions
  ($\diamo$) at an equivalent quark mass.}
\vspace{-0.2cm}
\end{centering}
\end{figure}

\section{TEST}

The rotation (\ref{transf}) is mathematically equivalent to the 
transformation of the mass:
\begin{equation}
m \rightarrow m^\phi = m \,(\cos 2\phi - \mbox{i} \gamma_5 \sin 2\phi).
\end{equation}
In the following we shall use this transformation. Furthermore we take
$\phi = \pi/4$. But we found that the method works equally well
for angles as small as $|\phi| = 0.05$.

To test the idea, we have looked at exceptional configurations which
we encountered in our runs with improved fermions~\cite{Gockeler}
on the $16^3 32$ lattice at $\beta = 5.7$. If the method works at this
coupling, we expect that it will work also at larger values of
$\beta$. In Fig.~3 we show the pion correlator for a single
exceptional configuration before and after the rotation. Before the
rotation we find a similar peak in the pion correlator as seen in
Fig.~1. After the rotation we find a correlator which is perfectly 
well behaved. It turns out to be in good agreement with the ensemble 
average~\cite{Gockeler}. Note that in first
approximation the physical quark mass is left unchanged by the rotation.
For comparison we have computed the pion correlator also for Wilson
fermions. The hopping parameter was chosen so as to give roughly the same
pion mass. We find good agreement with the rotated action. 

\section{DISCUSSION}
\label{discussion}

From both, Figs.~1 and 3 it follows that the (suspected) near-zero 
eigenmode of ${\cal K}$ acts as a source term, and that the excitation
it induces decays
exponentially with roughly the mass of the pion. We would not expect
to find such a behavior for an instanton configuration. There is also
no sign of an abnormal fluctuation found in the case of Wilson
fermions. 

We conclude that the problem of exceptional configurations may be
resolved by a chiral rotation of the critical Wilson-Dirac
operator. The method allows calculations at any quark mass. On top of
that, the results of the improvement program remain valid.

\end{document}